\documentclass[twocolumn,showpacs]{revtex4}
\usepackage{psfig}
\begin{document}
\title{Omega photoproduction beyond the resonance region  at small u}
\author{A. Sibirtsev$^1$, K. Tsushima$^{1,2,3}$ and S. Krewald$^{1,2}$}
\affiliation{$^1$Institut f\"ur Kernphysik, Forschungszentrum J\"ulich,
D-52425 J\"ulich \\
$^{2}$Special Research Center for the Subatomic 
Structure of Matter (CSSM)  and Department of 
Physics and Mathematical Physics, 
University of Adelaide, SA 5005, Australia \\
$^3$ Department of Physics and Astronomy, University
of Georgia, Athens, GA 30602 USA \\ }

\begin{abstract}
Within the meson exchange model we study $\omega$ meson 
photoproduction at energies above the $s$ channel resonance region. 
Different model prescriptions for the $\omega{NN}$
vertex function are investigated imposing gauge invariance as well as
crossing symmetry.  The calculations 
reproduce the energy dependence of the differential $\omega$ photoproduction 
cross sections at moderate $|u|$ for $E_\gamma{\le}4.7$~GeV, which 
previously were discussed as an indication of the hard interaction 
between the photon and quarks of the nucleon. 
\end{abstract}

\maketitle
\section{Introduction}
At energies  beyond the $s$ channel resonance region the  
vector meson photoproduction off the nucleon is traditionally discussed 
in terms of exchanges in the $t$ and $u$ channels. 

At forward angles the vector meson photoproduction is dominated by 
the exchanges in the $t$ channel.
 A long time ago, Berman and Drell~\cite{Berman} proposed 
that at low energies and small $t$
meson photoproduction can be well understood in terms of the 
exchanges of light mesons. 

Most recently,  Donnachie and 
Landshoff~\cite{Donnachie1,Donnachie2} and Laget~\cite{Laget1}
illustrated that at high energies the photoproduction  can be well 
described by  soft pomeron and meson Regge trajectories. 

Furthermore, the high energy data~\cite{Breitweg} on $\rho$ 
photoproduction at $|t|$ above 
$\simeq$0.4~GeV$^2$ require some additional contribution,  which might 
stem from the hard pomeron exchange~\cite{Donnachie3}.
The data~\cite{Adloff} on $J/\Psi$ meson photoproduction
 can be  reproduced well by an additional introduction of the hard 
pomeron exchange in the $t$ channel
~\cite{Donnachie3,Sibirtsev1,Donnachie4} even at small $|t|$.

At backward angles, i.e. where $|t|$ becomes large and $|u|$ 
still remains small, the photoproduction is dominated by exchanges in 
the $u$ channel. At low energies beyond the resonance region,
the production amplitude is given by the nucleon exchange.
At high energies  it is due to the contribution from  
the nucleon Regge trajectory.

Meson photoproduction at the kinematical region
 where both $t$ and $u$ are 
large can be dominated by hard interactions between the photon and the 
quarks of the nucleon. It is believed that this region is beyond the 
applicability of the standard boson exchange and Regge models. Moreover,
a typical signature~\cite{Brodsky} of hard processes is a strong 
${\propto}s^{-8}$ dependence of the differential photoproduction 
cross section.

\begin{figure}[h]
\vspace*{-6mm}
\hspace*{-2mm}\psfig{file=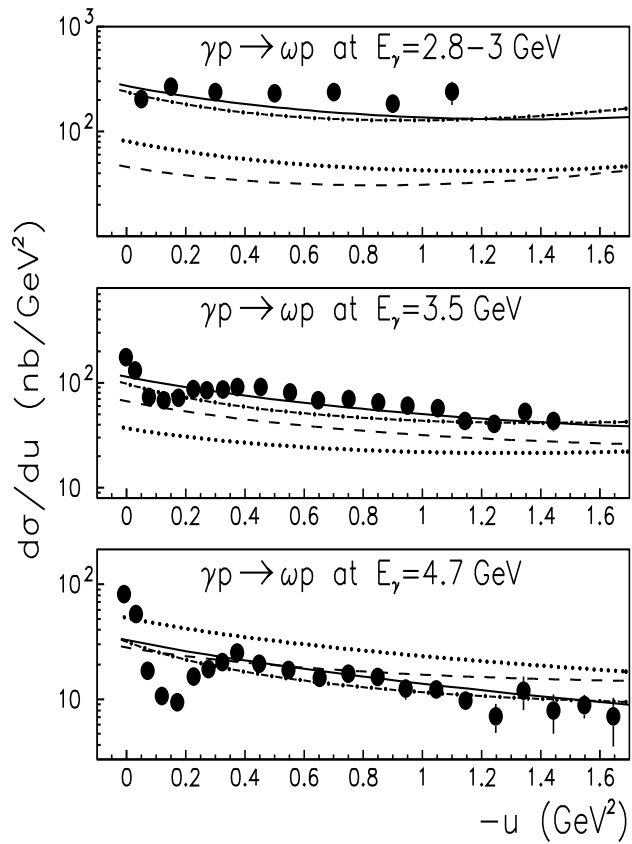,width=9.5cm,height=12.5cm}
\vspace*{-9mm}
\caption[]{The data~\cite{Clifft} on the differential  
$\gamma{p}{\to}\omega{p}$ cross section as function of $u$ for 
different photon energies $E_\gamma$. The lines show our 
meson exchange calculations with different prescriptions for the 
$\omega{NN}$ vertex functions given by models A (dashed), 
B (dotted), C(dot-dashed), and D (solid).} 
\label{gaomk2}
\end{figure}

Experimental results~\cite{Clifft} on backward $\omega$ meson photoproduction
at photon energies $E_\gamma$ between 2.8 and 4.8~GeV  
collected at Daresbury Laboratory indicate
a prominent structure shown in  Fig.~\ref{gaomk2}. The differential
$d\sigma{/}du$ cross section has a dip around $u{\simeq}0.15$~GeV$^2$,
which becomes more pronounced with increasing the photon energy.

It was proposed~\cite{Clifft} to parameterize the $\gamma{p}{\to}\omega{p}$ 
differential cross section as a sum of contributions from 
the Regge trajectory exchange and hard mechanism. Indeed the energy 
dependence of the $\omega$ meson photoproduction cross section 
$d\sigma{/}du$ at different values of $u$  was fitted well by 
\begin{equation}
\frac{d\sigma}{du}= |a(u)s^{\alpha(u)-1}+b(u)\exp^{i\phi(u)}
s^{-n/2}|^2,
\end{equation}
with fixed power $n{=}8$ and parameters $a$, $b$ and $\phi$ adjusted
to the data~\cite{Clifft}. It was found that $\alpha(u)$ 
corresponds to the nucleon trajectory, that the parameter $b$
does not depend on $u$ and the phase $\phi{=}90^o$ indicating that
the Regge exchange and hard mechanism do not interfere.

Finally, the data~\cite{Clifft} on backward $\omega$ photoproduction 
off a nucleon at $E_\gamma$ between 2.8 and 4.8~GeV were interpreted
as due to the dominant contribution from the nucleon Regge trajectory 
exchange at $|u|{\le}$0.2~GeV$^2$ and an additional  contribution from
the hard interaction between the photon and the quarks of the target 
nucleon.  

On  the other hand, the data~\cite{Clifft1} on backward $\rho$ and $f$ 
meson photoproduction collected at the same spectrometer and for
 the same
range of photon energies do not indicate a dip around
$u{\simeq}{-}0.15$~GeV$^2$ as well as a  strong $s^{-8}$ dependence.

Furthermore high precision measurements~\cite{Anderson} at SLAC 
of the backward photoproduction of  charged pions  off a nucleon at 
photon energies between 4.1 and 14.8~GeV and $-1.8{\le}u{\le}0.05$~GeV$^2$ 
also do not indicate a dip at $u{\simeq}$--0.15~GeV$^2$.
SLAC data~\cite{Tompkins} on backward $\pi^o$ photoproduction 
at photon energies $E_\gamma$=6, 8, 12 and 18~GeV show
a  $s^{-3}$ dependence
with no sign of a dip at $u$ between -1 and 0~GeV$^2$.

Therefore, among available  data on meson photoproduction at large angles
only the $\gamma{p}{\to}\omega{p}$ reaction shows a very specific
signature which might be interpreted as an indication of hard interactions
between the photon and the nucleon. However it is quite difficult to
motivate that the $\omega$ photoproduction can be sensitive to 
hard processes whereas photoproduction of $\pi$, $\rho$ and $f$
mesons are not, since hard interactions are driven by the partonic 
structure of the hadrons. 

Here we investigate  whether the data~\cite{Clifft}  on backward $\omega$ 
meson photoproduction might be understood in terms of a standard hadronic 
approach as a meson exchange model. At photon energies above
the $s$ resonance region we adopt the $\pi$, $\eta$ and $\sigma$ 
exchanges in the $t$ channel as well as the nucleon exchange in the $s$ and 
$u$ channels. Within meson exchange models, the contribution from 
the nucleon exchange dominates at large angles. 

It is important to note that 
within the meson exchange model, the  difference between the data on backward 
$\omega$ and  $\pi$, $\rho$ and $f$ meson photoproduction 
can be reasonably motivated, since the couplings and the form factor
functions in the meson-nucleon-nucleon vertices are not 
necessarily identical for the  various mesons. 
Moreover, theoretical analysis~\cite{Machleidt1,Machleidt2} 
of the nucleon-nucleon interaction really result in  different 
vertex functions for different mesons.

\section{The model}
The  vector meson photoproduction at small four momentum $t$ transfer  
has been studied theoretically for a long time~\cite{Drell,Alles,Joos}. 
Considering the contributions from all possible intermediate states that
are lighter  than the  produced vector mesons, Joos and 
Kramer~\cite{Joos} formulated the $\pi$, $\eta$ and 
$\sigma$ exchange model. 

The coupling constants for the $\pi$, $\eta$ and $\sigma$ 
exchanges were fixed either from the analysis of nucleon-nucleon 
scattering or from the relevant  partial decay widths  of the vector 
meson to the photon and the pseudoscalar meson. The  $\omega\gamma\sigma$
coupling was not  specified explicitly. 
Form factors  in the interaction vertices were
not taken into account~\cite{Gottfried,Watson}. 

However, the model in its original formulation~\cite{Drell,Alles,Berman,Joos} 
was not able to reproduce the data 
on $\omega$ photoproduction at energies 1.8${\le}E_\gamma{\le}$5.8~GeV
collected later~\cite{Aachen1,Ballam1,Ballam2,Davier}. 
We notice that
the $\omega$ photoproduction data~\cite{Drell,Alles,Berman,Joos} at
low $|t|$ can be well described by the model of Ref.~\cite{Joos}
when introducing  the relevant form factors at the
interaction vertices.

Following the model of Refs.~\cite{Drell,Alles,Berman,Joos} and
evaluating the interaction vertices by vector meson dominance
Friman and Soyeur~\cite{Friman} reproduced the data~\cite{LB1} 
on total $\gamma{p}{\to}\omega{p}$ cross section at photon
energies below 2~GeV and differential $d\sigma{/}dt$ 
omega photoproduction cross sections~\cite{Aachen1} at 
$|t|{\le}$0.5~GeV$^2$ for 1.4${\le}E_\gamma{\le}$1.8~GeV. 
In this model~\cite{Friman} the photon couples to the $\rho$ and the $\omega$
and the $\rho{N}{\to}\omega{N}$
transition is given by $\pi$ meson exchange, while the
$\omega{N}{\to}\omega{N}$ scattering was described by $\sigma$
exchange. Furthermore, the  interaction vertices were dressed
by form factors and unknown parameters for the $\sigma$ meson 
exchange were adjusted to the data~\cite{Aachen1}.

Obviously, since both  models~\cite{Joos,Friman} discussed 
above account only for the $t$-channel contribution, they are not able to  
reproduce the data~\cite{Clifft} collected at small $|u|$.

Very recently Oh, Titov and T.S.H. Lee~\cite{Oh1} investigated
$\omega$ photoproduction in the resonance region by considering
the $\pi$, $\eta$ and pomeron exchanges in the  $t$ channel
and the excitation of the nucleon and baryonic resonances in the 
$s$ and $u$ channels. The form factors and coupling constants
for $\pi$, $\eta$ and nucleon exchanges were adopted from the 
literature. The parameters for the  pomeron exchange were  
fixed~\cite{Donnachie1,Donnachie5,Donnachie6} at high energies. 
Furthermore, the resonance parameters were fitted to the 
$\gamma{p}{\to}\omega{p}$ differential cross sections~\cite{Klein} 
collected at photon energies 1.23${\le}E_\gamma{\le}$1.92~GeV.
The calculations~\cite{Oh1} also well reproduce the 
data~\cite{Ballam3} at small $|t|$ for $E_\gamma$=2.8 and 4.7~GeV.
However, the model could not reproduce the $\omega$ photoproduction
data~\cite{Clifft} at small $|u|$.
 
\begin{figure}[h]
\vspace*{-5mm}
\hspace*{-5mm}\psfig{file=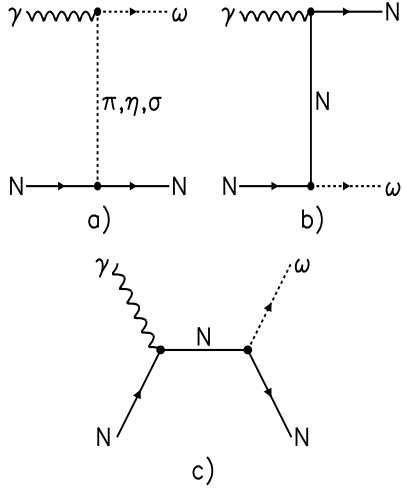,width=7.cm,height=8.cm}
\vspace*{-8mm}\caption[]{$\omega$ meson photoproduction mechanism in $t$ (a),
$u$ (b) and $s$ (c) channels.}
\label{gaomk6}
\end{figure}

Here we consider $\omega$ meson photoproduction due to the pseudoscalar
$\pi$ and $\eta$ and scalar $\sigma$ meson exchanges and the direct 
and crossed nucleon terms. The relevant diagrams are shown in the
Fig.~\ref{gaomk6}. Furthermore, the squared invariant collision energy
$s$ and the squared  $t$ and $u$ four momenta transfers are given by
\begin{eqnarray}
s= (p+k)^2=(p^\prime +q)^2, \\
t=(q-k)^2= (p^\prime -p)^2, \\ 
u=(q-p)^2=(p^\prime -k)^2, 
\end{eqnarray}
where $k$, $q$, $p$, and $p^\prime$ are the  four momenta
of the photon, $\omega$ meson, initial nucleon and final nucleon,
respectively.  

The effective Lagrangian densities used for the 
evaluation of the $\pi$ and $\eta$ meson exchange amplitudes 
are given as
\begin{eqnarray}
{\cal L}_{\omega\gamma \varphi} &=&
\frac{e g_{\omega\gamma \varphi}}{m_\omega }\,
\epsilon^{\mu\nu\alpha\beta}
\partial_\mu \omega_\nu \partial_\alpha A_\beta\, \varphi,\qquad
\nonumber\\
{\cal L}_{\varphi NN} &=&
-i g_{\varphi NN} \bar N \gamma_5 N \varphi,
\end{eqnarray}
where $\varphi$ denotes the $\pi^0$ or the $\eta$ meson, while
$A_\beta$ is the photon field. Furthermore we adopt the $\pi{NN}$ 
and $\eta{NN}$  coupling constants as $g_{\pi NN}{=}13.26$ and 
$g_{\eta NN}{=}3.53$, respectively. The $g_{\omega\gamma\pi}$=1.82 and
$g_{\omega\gamma\eta}$=0.42 coupling constants were evaluated from the
partial $\omega{\to}\gamma\pi$ and  $\omega{\to}\gamma\eta$ 
decay widths~\cite{PDG}, respectively. 

Finally, the $\gamma{p}{\to}\omega{p}$ reaction amplitude due to the 
$\pi$ and $\eta$ meson exchange is given as
\begin{eqnarray}
{\cal M}_{\pi,\eta} =\sum_{\varphi = \pi,\eta}
\frac{-ie g_{\omega\gamma\varphi} g_{\varphi NN}}{m_\omega}\,
\frac{1}{t-m_\varphi^2}
\nonumber \\ \times
\varepsilon^{\mu\nu\alpha\beta} q_\mu \varepsilon^*_\nu(q) k_\alpha
\varepsilon_\beta(k) \, \bar{u}(p^\prime) \gamma_5 u(p).
\end{eqnarray}

Moreover, at each interaction 
vertex we introduce the phenomenological form factor
\begin{equation} 
F (t) = \frac{\Lambda^2 - m_{ex}^2}{\Lambda^2 -t},
\label{ff1}
\end{equation}
where $m_{ex}$ is the mass of the exchanged meson and the
cutoff parameters in the $\pi{NN}$ and $\eta{NN}$ vertices
were fixed as $\Lambda_{\pi{NN}}$=0.7~GeV and $\Lambda_{\eta{NN}}$=1~GeV.
The cutoff parameters of the form factors in the $\omega\gamma\pi$ 
and  $\omega\gamma\eta$ vertices were fitted to the data 
as will be specified later.

The  $\sigma$ meson exchange amplitude was calculated from the 
following effective Lagrangians:
\begin{eqnarray}
{\cal L}_{\omega\gamma\sigma} &=& e m_\omega g_{\omega\gamma\sigma}
\omega_\beta A^\beta\,\sigma \nonumber\\
{\cal L}_{\sigma NN} &=& g_{\sigma NN}\bar N  N \sigma,
\end{eqnarray} 
where the $\sigma{NN}$ coupling constant was taken as
$g_{\sigma NN}$=8, while the $\omega\gamma\sigma$ coupling constant 
was fitted to the data. 

The photoproduction amplitude due to the $\sigma$ meson
exchange is then naively given as 
\begin{equation}
{\cal M}_\sigma = \frac{e m_\omega 
g_{\omega\gamma\sigma}g_{\sigma NN}}{t-m_\sigma^2}
\, \varepsilon_\mu^*(q)\varepsilon^\mu(k)
\bar{u}(p^\prime) u(p). 
\end{equation}

Again we introduce  monopole form factor of Eq.~(\ref{ff1}) in 
the interaction vertices with $\Lambda_{\sigma{NN}}$=1~GeV and 
adjust the cutoff $\Lambda_{\omega\gamma\sigma}$ to the data. 
Furthermore, the gauge invariant $\sigma$ exchange 
amplitudes is now given as
\begin{eqnarray}
{\cal M}_\sigma &=& \frac{e m_\omega
g_{\omega\gamma\sigma}g_{\sigma NN}}{t-m_\sigma^2}\nonumber\\
&\times& \varepsilon_\mu^*(q)\, (g^{\mu\nu}{-}\frac{k^\mu q^\nu}{q \cdot k}) 
\,\varepsilon_\nu(k)
\bar{u}(p^\prime) u(p).
\end{eqnarray}

The direct and crossed nucleon exchange amplitudes
due to the diagrams depicted in Fig.~\ref{gaomk6}c) and 
Fig.~\ref{gaomk6}b), respectively, were
evaluated from the following interaction Lagrangians:
\begin{eqnarray}
{\cal L}_{\gamma NN} & = &
- e \bar{N} \left( \gamma_\mu \frac{1+\tau_3}{2} {A}^\mu
- \frac{\kappa_p^{}}{2M_N^{}} \sigma^{\mu\nu} \partial_\nu A_\mu \right) N,
\nonumber\\
{\cal L}_{\omega NN} & = &
- g_{\omega NN}^{} \bar{N} \left( \gamma_\mu {\omega}^\mu
- \frac{\kappa_\omega}{2M_N^{}} \sigma^{\mu\nu}
\partial_\nu \omega_\mu \right) N,
\end{eqnarray}
where the anomalous magnetic moment of the proton 
was taken as $\kappa_{p}$=1.79, while the $\omega{NN}$ coupling 
constant was fitted to the data. Moreover, in the following 
calculations we adopt $\kappa_\omega$=0.

Ignoring the form factors for the $\gamma{p}{\to}\omega{p}$ reaction, 
the reaction amplitudes due to the nucleon exchange in the $s$ and 
$u$ channels are given as
\begin{eqnarray}
{\cal M}_N={e g_{\omega NN}}\,
{\bar u}(p^\prime)\, \varepsilon^*_\mu(q)\gamma^\mu
\frac{{\not\!p}{+}{\not\!k}{+}m_N}{s-m_N^2}\nonumber \\ 
\times [\gamma_\nu+\frac{i\kappa_p}{2m_N}
\sigma_{\nu\alpha} k^\alpha] \, \varepsilon^\nu(k)u(p), \\
\tilde{{\cal M}}_N ={e g_{\omega NN}}\,
{\bar u}(p^\prime)\, \varepsilon^*_\mu(q)
[\gamma_\nu+\frac{i\kappa_p}{2m_N}
\sigma_{\nu\alpha} k^\alpha]\nonumber \\ 
\times  
\frac{{\not\!p}{-}{\not\!q}{+}m_N}{u-m_N^2}\,
\gamma^\mu \varepsilon^\nu(k)u(p).
\end{eqnarray}
The form factor in the $\omega{NN}$ vertex will be specified in
the next section. 

Finally, the model parameters which we fixed 
prior to the global fitting to the data are listed in 
Table~\ref{tab1}. However let us to remind that for each
meson exchange diagram shown in the Fig.\ref{gaomk6}a)
the form factors in the $\omega\gamma\pi$, 
$\omega\gamma\eta$ and  $\omega\gamma\sigma$ vertices 
were adjusted to the data, while those in the $\pi{NN}$, 
$\eta{NN}$ and $\sigma{NN}$vertices, respectively, were fixed.
Since the ${\cal M}_\pi$, ${\cal M}_\eta$ and
${\cal M}_\sigma$ amplitudes contain the product of the form factors
in both vertices, the model could not be used for a
unique determination of the relevant vertex function 
parameters.     

\begin{table}[h]
\caption{Coupling constants $g$ and  cut-off parameters $\Lambda$
fixed in our calculations. The cut-off parameters are given for 
the form factor of Eq.~(\protect\ref{ff1}).}
\label{tab1}
\vspace{0.5mm}
\begin{ruledtabular}
\begin{tabular}{ccc}
 \hspace{5mm} vertex \hspace{5mm} &  \hspace{5mm}$g$ \hspace{5mm} & 
 \hspace{5mm} $\Lambda$ (GeV) \hspace{5mm} 
\vspace{0.6mm} \\
\colrule
$\pi{NN}$ & 13.26  & 0.7  \\
$\eta{NN}$ & 3.53 & 1.0 \\
$\sigma{NN}$ & 8.0 & 1.0 \\
$\omega\gamma\pi$ & 1.82 & -- \\ 
$\omega\gamma\eta$ & 0.42 & -- \\
\end{tabular}
\end{ruledtabular}
\end{table}

\section{The $\omega{NN}$ vertex function}
A naive evaluation of the amplitude and/or 
the introduction of phenomenological form factors in 
the photoproduction reactions at the level of Born amplitudes 
violates gauge invariance. There are a few common 
recipes~\cite{Ohta,Workman,Haberzettl1,Haberzettl2,Davidson}
proposed to restore gauge invariance. The dominant contribution
at small $u$ comes from the nucleon exchange current. Therefore, the data
on $\omega$ photoproduction at small $u$ should be
sensitive to the prescription of the $\omega{NN}$ vertex function
as well as to the methods for restoration of gauge invariance of the 
nucleon exchange amplitude.

In order to understand whether it is necessary to use 
form factors in the  $\omega{NN}$ vertices at $s$ and $u$ 
channels of the diagrams (b) and (c) in Fig.\ref{gaomk6} we 
fit the data  without $\omega{NN}$ form factors
and vary the $g_{\omega{NN}}$ and $g_{\omega\gamma\sigma}$
coupling constants and cut-off parameters at the $\omega\gamma\pi$,
$\omega\gamma\eta$ and $\omega\gamma\sigma$ vertices. We denote 
these  calculations as Model A. 

Furthermore, we use simultaneously the $\gamma{p}{\to}\omega{p}$ data on 
the $d\sigma{/}du$~\cite{Clifft} and the $d\sigma{/}dt$  differential 
cross sections available at photon energies 2.8${\le}E_\gamma{\le}$4.7~GeV.
The global fit was performed with Minuit~\cite{James}
and includes 120 experimental points.

\begin{table}[h]\vspace{-2mm}
\caption{Model parameters fitted to the data~\cite{Clifft}.
Model A is the calculations without form factor at the $\omega{NN}$
vertex. Model B stands for the calculations with form factor
of Eq.~(\ref{ff2}) at the $\omega{NN}$ vertex. Model~C denotes 
the calculations with the form factor of Eq.~(\ref{ff3}) at 
the $\omega{NN}$ vertex. Model~D denotes 
the calculations with the form factor of Eq.~(\ref{ff31}). 
The cut-off parameters are given in GeV. 
The $\chi^2$ is total chi-square.}
\label{tab2}
\vspace{0.8mm}
\begin{ruledtabular}
\begin{tabular}{ccccc}
\hspace{0mm}parameter\hspace{1mm} & \hspace{1mm}model A\hspace{2mm} & 
\hspace{2mm}model B\hspace{2mm} & \hspace{2mm}model C\hspace{2mm}  &
\hspace{2mm}model D\hspace{2mm} 
\vspace{0.6mm} 
\\
\colrule
$\Lambda_{\omega\gamma\pi}$ &0.10$\pm$0.07 &0.10$\pm$0.04 & 
0.40$\pm$0.01&0.60$\pm$0.03 \\
$\Lambda_{\omega\gamma\eta}$ &0.1$\pm$0.3 &0.1$\pm$0.6 & 
0.4$\pm$0.1&1.4$\pm$0.9\\
$g_{\omega\gamma\sigma}$ & 7.2$\pm$0.2 &10$\pm$1 & 
1.4$\pm$0.2&1.11$\pm$0.01\\
$\Lambda_{\omega\gamma\sigma}$ & 0.59$\pm$0.01 &
0.52$\pm$0.01 & 0.83$\pm$0.07&1.0$\pm$0.1\\
$g_{\omega{NN}}$ &0.22$\pm$0.01 & 0.26$\pm$0.02 & 
5.8$\pm$0.5&1.7$\pm$0.1 \\
$\Lambda_{\omega{NN}}$ & --- & 2.21$\pm$0.01& 0.30$\pm$0.01&1.85$\pm$0.03 \\
$\tilde{\Lambda}_{\omega{NN}}$ & --- & --- & 1.23$\pm$0.03&---\\
$\chi^2$ &1796 & 1799  & 614 & 598\\
\end{tabular}
\end{ruledtabular}
\end{table}

\begin{figure}[b]
\vspace{-6mm}
\hspace*{-1mm}\psfig{file=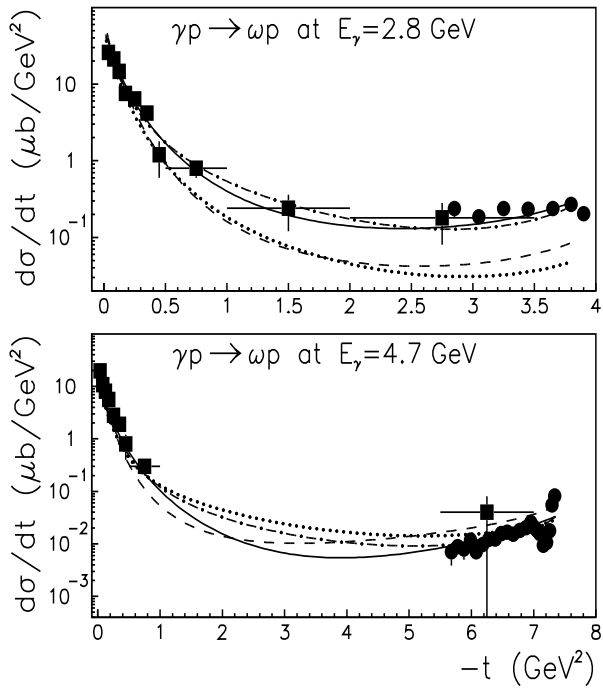,width=9.cm,height=10.4cm}\vspace{-6mm}
\caption[]{The data~\cite{Clifft,Ballam3} on the  differential  
$\gamma{p}{\to}\omega{p}$ cross section as a function of $t$ for 
different photon energies $E_\gamma$. The lines show our 
calculations with different prescriptions for the 
$\omega{NN}$ vertex functions given by model A (dashed), 
B (dotted), C (dashed-dotted), and D (solid).}
\label{gaomk1}
\end{figure}

The results obtained with model~A are shown by the dashed lines 
in Fig.~\ref{gaomk2} and Fig.~\ref{gaomk1}. The relevant sets of 
model parameters and the total $\chi^2$ are listed in Table~\ref{tab2}. 
The calculations  describe the  data at small
$|u|$ at $E_\gamma$=4.7~GeV well. However,  model~A  could not 
reproduce the dependence of the $\omega$ photoproduction differential 
cross section $d\sigma{/}du$ on the photon energy. Moreover, the fit
results in large $\omega\gamma\sigma$ and very small
$\omega{NN}$ coupling constants. Therefore we
consider  the parameters of model~A as unreasonable. 
Finally we conclude that the calculations without a form factor at the
$\omega{NN}$ vertex cannot not reproduce the complete set of data on the
$\gamma{p}{\to}\omega{p}$ reaction.

Now we have followed Refs.~\cite{Oh1,Haberzettl2} and included a 
form factor at the $\omega{NN}$ vertex  in the
$u$ and $s$ channels
\begin{equation}
F(r) =\frac {\Lambda^4}{\Lambda^4+(r-m_N^2)^2},
\label{ff2}
\end{equation}
where $r$=$s$ or $u$. The cut-off parameter $\Lambda$ was
fitted to the data.

Furthermore, the gauge invariant  nucleon exchange 
amplitudes are now given as
\begin{eqnarray}
{\cal M}_N &=& {e g_{\omega NN}}\,
{\bar u}(p^\prime)\, \varepsilon^*_\mu(q)[\gamma^\mu{-}k^\mu
\frac{\not\!q}{q{\cdot}k}]
\frac{{\not\!p}{+}{\not\!k}{+}m_N}{s-m_N^2}\nonumber \\ 
&\times& F(s)\,[\gamma_\nu{+}\frac{\kappa_p}{2m_N}
\gamma_\nu{\not\!k}{-}\frac{q_\nu\not\!k}{q{\cdot}k}] \, 
\varepsilon^\nu(k)u(p), \\
\tilde{{\cal M}}_N &=& {e g_{\omega NN}}\,
{\bar u}(p^\prime)\, \varepsilon^*_\mu(q)
[\gamma_\nu{+}\frac{\kappa_p}{2m_N}
\gamma_\nu{\not\!k}{-}\frac{q_\nu\not\!k}{q{\cdot}k}]\nonumber \\ 
&\times&  F(u)\,
\frac{{\not\!p}{-}{\not\!q}{+}m_N}{u-m_N^2}\,
[\gamma^\mu{-}k^\mu\frac{\not\!q}{q{\cdot}k}] \varepsilon^\nu(k)u(p).
\end{eqnarray}

We denote the calculations with the form factor of Eq.~(\ref{ff2}) 
as model~B and show the results by the dotted lines in Fig.~\ref{gaomk2} 
and Fig.~\ref{gaomk1}. The parameters of  model~B are given in 
Table~\ref{tab2}.

The calculations by model~B again could not reproduce the 
dependence of the $\gamma{p}{\to}\omega{p}$ differential 
cross section on the photon energy.  The data on 
$d\sigma{/}dt$ and $d\sigma{/}du$ at $E_\gamma$=4.7~GeV
are well described, however.  We also notice that our results shown in 
Fig.~\ref{gaomk2} are different from those reported in 
Ref.~\cite{Oh1}. We attribute this discrepancy to the fact that 
in our study the complete set of available $\omega$ photoproduction
data on both $d\sigma{/}dt$ and $d\sigma{/}du$ is included for the 
whole energy range
2.8${\le}E_\gamma{\le}$4.7~GeV, while Oh, Titov and T.S.H. 
Lee~\cite{Oh1} adjusted the model parameters to the $d\sigma{/}dt$
data only. 

\begin{figure}[h]
\vspace{-5mm}
\hspace*{-1mm}\psfig{file=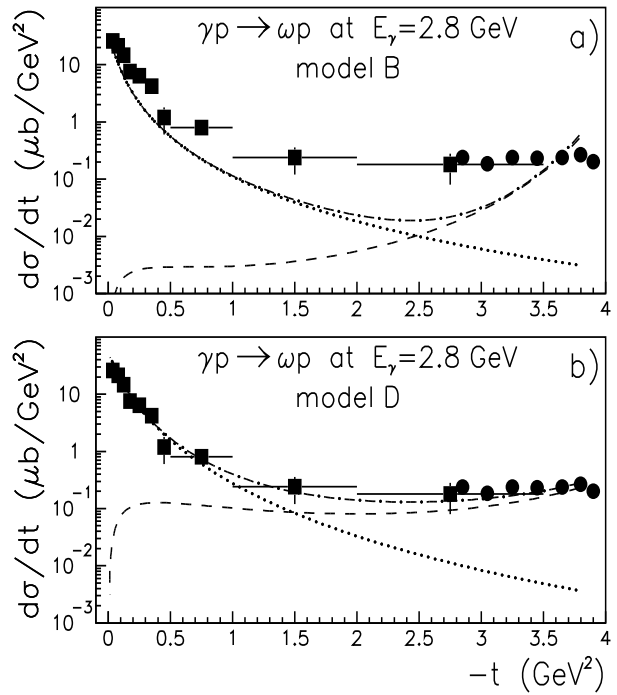,width=9.1cm,height=10.5cm}\vspace{-5mm}
\caption[]{The differential $\gamma{p}{\to}\omega{p}$ cross section as a
function of $t$ for $E_\gamma$=2.8~GeV. The data are from 
Refs.~\cite{Clifft,Ballam3}. The calculations were done by model~B with 
parameters from Ref.~\cite{Oh1} (a) and by model~D (b). Dashed lines:
contributions from the nucleon exchange; dotted lines:  meson exchange;
dashed-dotted lines: both nucleonic and meson
exchanges.}
\label{gaomk7}
\end{figure}

In order to solve the so-called A2-problem of pion photoproduction,
 Davidson and Workman~\cite{Davidson} recently suggested a new kind 
of form factor motivated by gauge invariance and crossing symmetry 
which reads as follows:
\begin{equation}
F(u,s)=F(u,\Lambda)+F(s,\tilde{\Lambda})-F(u,\Lambda)F(s,\tilde{\Lambda}),
\label{ff3}
\end{equation}
with the form factors taken in the form of Eq.~(\ref{ff2}).
The cut-off parameters $\Lambda$ and $\tilde{\Lambda}$ should be 
identical in order to satisfy crossing symmetry.
The calculations performed with this form factor are denoted as 
model~C. The results are displayed in
Fig.~\ref{gaomk2} and Fig.~\ref{gaomk1} by the dashed-dotted lines. 
The relevant parameters are listed in  Table~\ref{tab2}. It turned 
out that the crossing symmetry had to be broken, if a good fit to 
the data was required. The question arises
whether the idea of Davidson and Workman should be abandoned.

We found that a mild modification of the ansatz of Davidson and Workman
allows to maintain the crossing symmetry. We investigate the ansatz: 
\begin{equation}
F(u,s)=F(u,\Lambda)F(s,\Lambda).
\label{ff31}
\end{equation}
The form factor  of Eq.~(\ref{ff31}) has to multiply the sum of both 
the $s$ and $u$ channel diagrams. Model ~D refers to calculations performed 
with the prescription Eq.~(\ref{ff31}).

Model~D  describes  the $d\sigma{/}dt$ and $d\sigma{/}du$
data on $\omega$ meson photoproduction at 2.8${\le}E_\gamma{\le}$4.7~GeV 
better than all the other models( see e.g. the  $\chi^2$-values ) 
with the same number of parameters as is employed in model~B.
Moreover, we consider the parameters of  model~D as quite
reasonable. However, our calculations could not reproduce the
dip of the  $\gamma{p}{\to}\omega{p}$ differential cross section 
around $|u|{\simeq}$0.15~GeV$^2$ at the photon energy $E_\gamma$=4.7~GeV.
One might  speculate whether the appearance of a dip in $d\sigma{/}du$ 
needs additional experimental confirmation.

As is shown in  Fig.~\ref{gaomk1}, the four   different models  
reproduce the differential $\omega$ photoproduction cross
section at small $|t|$ at different photon energies quite 
well, but substantially differ at  $|t|{\ge}$1~GeV$^2$. Therefore 
we conclude that the 
prescription of the $\omega{NN}$ vertex function as well as the 
model parameters  substantially influence the results in 
the region of moderate and large $|t|$. 

To illustrate our finding explicitly, we calculate the
$\gamma{p}{\to}\omega{p}$ differential cross section $d\sigma{/}dt$
by  model~B with the set of parameters from Ref.~\cite{Oh1} taken as
\begin{eqnarray}
&g_{\pi NN}{=}13.26,  \Lambda_{\pi NN}{=}0.6, 
g_{\omega\gamma\pi}{=}1.823,   \Lambda_{\omega\gamma\pi}{=}0.9,&
\nonumber \\ 
&g_{\eta NN}{=}3.53,  \Lambda_{\eta NN}{=}1.0, 
g_{\omega\gamma\eta}{=}0.416, \Lambda_{\omega\gamma\eta}{=}0.9& 
\nonumber \\
&g_{\omega NN}{=}10.35, \Lambda_{\omega NN}{=}0.5,&
\label{para1}
\end{eqnarray}
where the cut-off parameters are given in GeV. Furthermore, in
Ref.~\cite{Oh1} the pomeron exchange was included instead of 
$\sigma$ meson exchange but it was shown that the contribution 
from the pomeron exchange is negligible at $E_\gamma$=2.8~GeV. Therefore
we also neglect the contribution from the $\sigma$ meson exchange in
reproducing the results from Ref.~\cite{Oh1} by model~B. 

Fig.~\ref{gaomk7}a) shows the differential $\gamma{p}{\to}\omega{p}$ 
cross section as a function of $t$ for $E_\gamma$=2.8~GeV. The
calculations are based on model~B with parameters given by Eq.~(\ref{para1}).
The dashed-dotted line shows the total contribution from 
both meson and nucleon exchanges,
while the dashed line indicates the contribution from nucleon
exchange only  and the dotted one the contribution from 
mesonic exchanges. The calculations
shown in Fig.~\ref{gaomk7}a) are identical to the ones reported in
Ref.~\cite{Oh1}. It is clear that model~B cannot reproduce the data
at $|t|{\ge}$0.5~GeV$^2$.

\begin{figure}[h]
\vspace{-6mm}
\hspace*{-1mm}\psfig{file=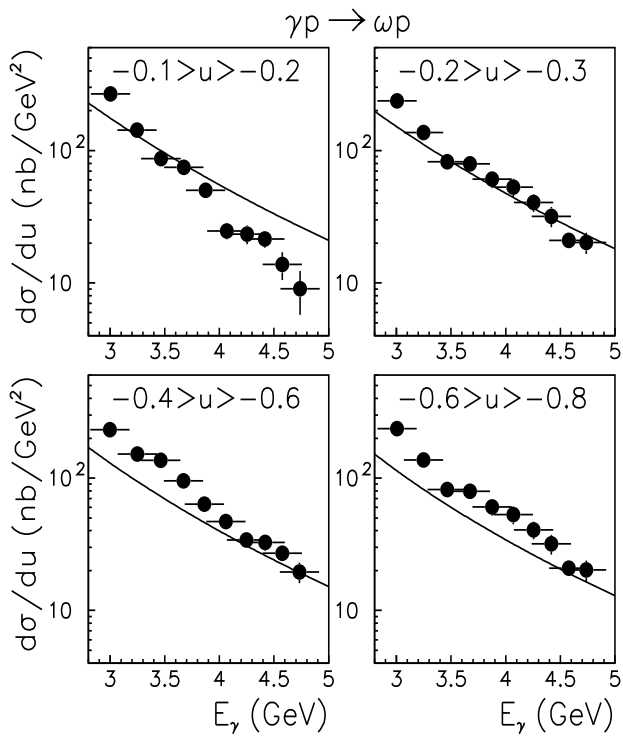,width=9.4cm,height=10.5cm}\vspace{-5mm}
\caption[]{The dependence of the differential $\gamma{p}{\to}\omega{p}$ 
cross section $d\sigma{/}du$ on the photon energy $E_\gamma$. The data
were taken  from Ref.~\cite{Clifft}. The solid line shows our calculations
based on  model~D.}
\label{gaomk3}
\end{figure}

Now, Fig.~\ref{gaomk7}b) shows that the calculations based on model~D 
reasonably reproduce the data~\cite{Clifft} for the whole range of $t$-values.
Comparing the two calculations one can easily notice 
that the contributions from
mesonic currents from model~B and model~C are almost identical,
whereas  the contributions from the nucleon exchange are substantially
different. Obviously, the prescription of the $\omega{NN}$ vertex is 
important in understanding the data on $\omega$ meson photoproduction
at large and moderate $|t|$.

\section{Energy dependence of the backward $\omega$ photoproduction}
Now we investigate the dependence of the differential 
$\gamma{p}{\to}\omega{p}$ cross section $d\sigma{/}du$ on the photon energy.
Let us to recall that the very steep $s^{-8}$ dependence~\cite{Brodsky} was
interpreted~\cite{Clifft} as an indication of  hard interactions between
the photon and the quarks of the nucleon.

\begin{figure}[h]
\vspace{-4mm}
\hspace*{-1mm}\psfig{file=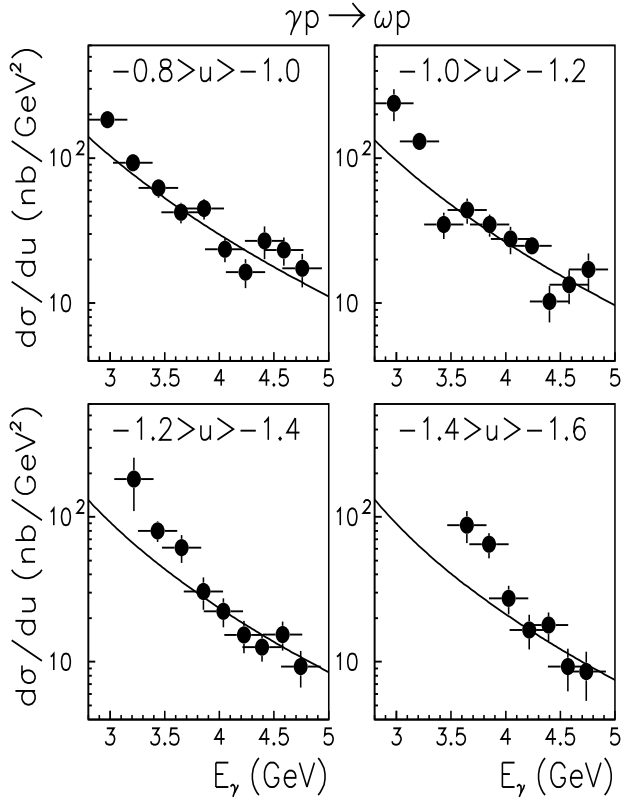,width=9.4cm,height=11.5cm}\vspace{-5mm}
\caption[]{The dependence of the differential $\gamma{p}{\to}\omega{p}$ 
cross section $d\sigma{/}du$ on the photon energy $E_\gamma$. The data
were taken  from Ref.~\cite{Clifft}. The solid line shows our calculations
based on model~D.}
\label{gaomk4}
\end{figure}

\begin{figure}[h]
\vspace{-6mm}
\hspace*{-1mm}\psfig{file=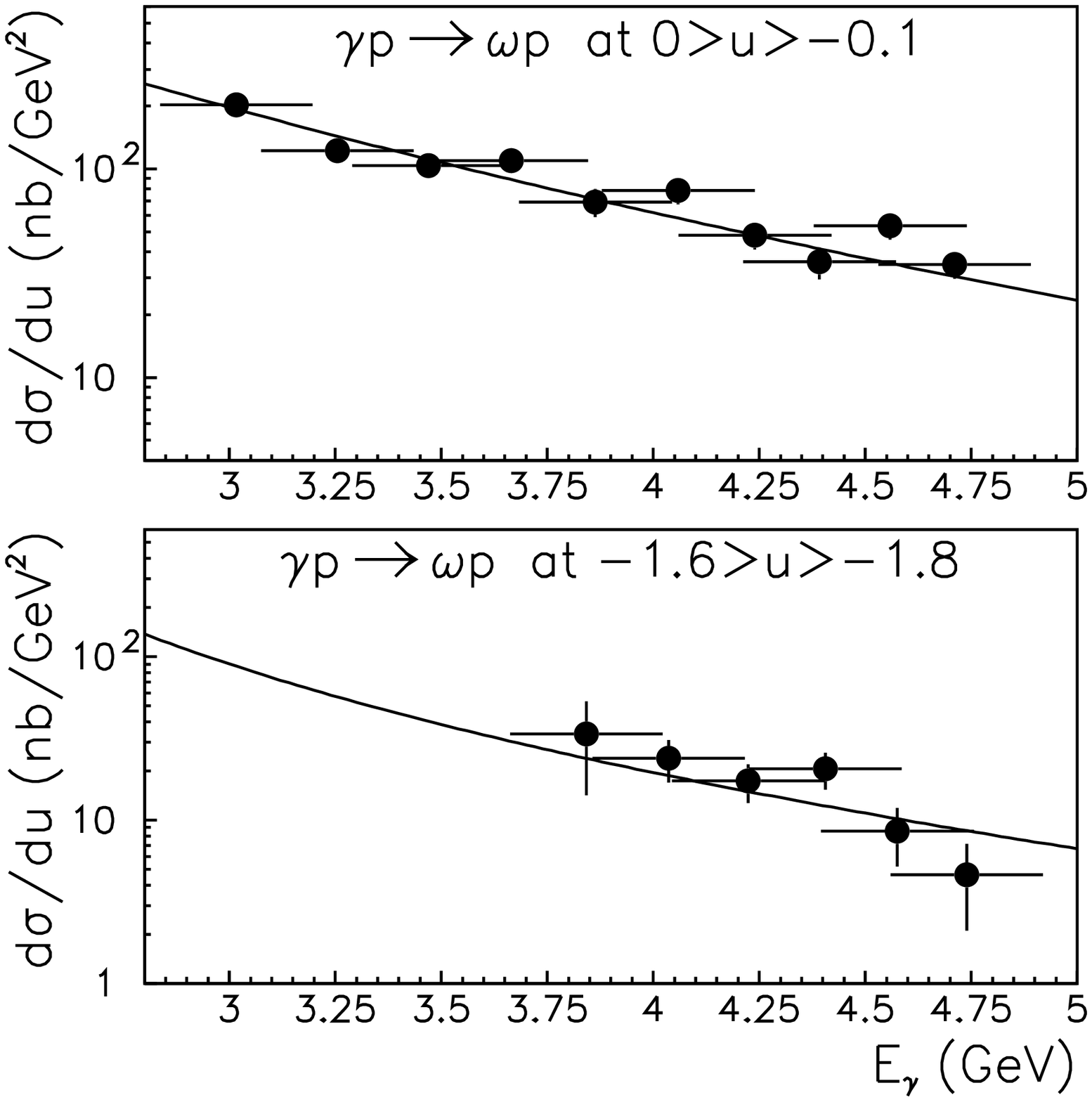,width=9.2cm,height=10.5cm}\vspace{-5mm}
\caption[]{The dependence of the differential $\gamma{p}{\to}\omega{p}$ 
cross section $d\sigma{/}du$ on the photon energy $E_\gamma$. The data
were taken  from Ref.~\cite{Clifft}. The solid line shows our calculations
based on  model~D.}
\label{gaomk5}
\end{figure}

Experimental data~\cite{Clifft} on the energy dependence of 
the $\omega$ photoproduction cross section $d\sigma{/}du$ are 
shown in Figs.\ref{gaomk3}-\ref{gaomk5}. The solid lines 
show our calculations based on model~D. The overall agreement between 
the data and the model calculations is reasonable, although there is a
systematical discrepancy at $-0.2{<}u{<}{-}0.1$~GeV$^2$ for photon energies
above 4~GeV, i.e. in the vicinity of the experimentally observed dip.
Furthermore, the calculations systematically underestimate the
part of the data at ${-}0.8{<}u{<}{-}0.4$~GeV$^2$, but rather well 
reproduce the energy dependence of the differential 
$\gamma{p}{\to}\omega{p}$ cross section $d\sigma{/}du$.

Two extreme situations are shown in  Fig.~\ref{gaomk5}. In 
Ref.~\cite{Clifft} the $\gamma{p}{\to}\omega{p}$ data at 
$-0.1{<}u{<}0$~GeV$^2$ were fitted by a 
$s^{-3.9{\pm}0.3}$ dependence, while experimental results
at $-1.8{<}u{<}{-}1.6$~GeV$^2$ were fitted by  a $s^{-9.3{\pm}2.1}$ 
dependence. Other data available for the range $-1.6{<}u{<}{-}0.1$~GeV$^2$
indicate that the power of $s$ is between the two extreme values given above.

Furthermore, our calculations can be fitted well by 
$d\sigma{/}du{\propto}s^{4.57}$ at
$-0.1{<}u{<}0$~GeV$^2$ and $d\sigma{/}du{\propto}s^{6.73}$ 
at $-1.8{<}u{<}{-}1.6$~GeV$^2$ respectively. Although the model does
not produce the very steep $s$ dependence, we conclude that the agreement
between the calculations and the experimental data is very reasonable.

\section{Summary}
The $\omega$ meson photoproduction for energies 2.8${\le}E_\gamma{\le}$4.7~GeV 
was studied within the boson exchange model. We consider
the $\pi$, $\eta$, $\sigma$ and nucleon exchanges. The mesonic 
exchanges dominate at small four momenta transfer squared $|t|$, 
while the nucleon exchange contributes at large $|t|$. 

It was found that the $\gamma{p}{\to}\omega{p}$ data~\cite{Clifft} 
available at large $|t|$ or small $|u|$ are very sensitive to the 
prescription of the $\omega{NN}$ vertex function. We investigated different 
recipes commonly used for restoration of gauge invariance in 
$\omega$ photoproduction. The inclusion of a phenomenological
form factor at the $\omega{NN}$ vertex which is consistent with
gauge invariance and crossing symmetry~\cite{Davidson} allows to
reproduce the data~\cite{Clifft} both at small $t$ and small
$u$.

However we found that within our approach it is impossible to
reproduce the dip in the differential  $\gamma{p}{\to}\omega{p}$
cross section $d\sigma{/}du$ around $u{\simeq}$--0.15~GeV$^2$.
Furthermore, since a similar dip was not observed in the backward
photoproduction of $\pi$, $\rho$ and $f$ mesons, we suggest that 
an additional experimental confirmation would be highly welcome.

We examined the energy dependence of the differential $\omega$
photoproduction cross section $d\sigma{/}du$. The experimentally
observed  steep ${\propto}s^{-8}$ dependence was interpreted ~\cite{Clifft}
as an indication~\cite{Brodsky} of hard interactions between the
photon and quarks of the nucleon. However, our meson-exchange calculations 
also quite reasonably reproduce the energy dependence of $d\sigma{/}du$.

\acknowledgements
Work of K.T. was supported by the Australian Research Council,  
and the Forschungzentrum J\"{u}lich. S.K. acknowledges support 
by grant N. 447AUS113/14/0 by the Deutsche Forschungsgemeinschaft.
Thanks are given to Profs. J.~Speth and A.W.~Thomas for their support 
of this work.

\end{document}